\documentclass[a4paper,fleqn]{cas-dc}

\usepackage{amsmath}
\usepackage{float}
\usepackage{placeins}
\usepackage[authoryear,longnamesfirst]{natbib}
\setcitestyle{numbers,square}

\usepackage{stfloats}

\def\tsc#1{\csdef{#1}{\textsc{\lowercase{#1}}\xspace}}
\tsc{WGM}
\tsc{QE}
\tsc{EP}
\tsc{PMS}
\tsc{BEC}
\tsc{DE}

\begin{document}
\let\WriteBookmarks\relax
\def\floatpagepagefraction{1}
\def\textpagefraction{.001}

\shorttitle{A High-Throughput Dark-Field Full-Field OCT System for Measuring Objects with Different Scattered Light Intensities}

\shortauthors{Y. Fan et~al.}

\title [mode = title]{A High-Throughput Dark-Field Full-Field OCT System for Measuring Objects with Different Scattered Light Intensities}                      
\tnotetext[1]{This work was financially supported by National Natural Science Foundation of China (No. 62375255).}

%
\author[1]{Youlong Fan}[style=chinese]






\affiliation[1]{organization={School of Physical Sciences},
    addressline={University of Science and Technology of China}, 
    city={Hefei},
    postcode={230041}, 
    country={China}}

\author[1]{Qingye Hu}[style=chinese]

\author[2]{Zhongping Wang}[style=chinese]

\author%
[2]
{Xiantao Wei}[style=chinese, orcid=0000-0003-1281-213X]
\cormark[1]
\ead{wxt@ustc.edu.cn}

\author[2]{Zengming Zhang}[style=chinese]

\affiliation[2]{organization={Physics Experiment Teaching Center},
    addressline={University of Science and Technology of China}, 
    city={Hefei},
    postcode={230041}, 
    country={China}}

\cortext[cor1]{Corresponding author}

\begin{abstract}
Based on high-throughput dark-field full-field optical coherence tomography, we designed and built an OCT system that can measure a variety of samples with different scattered light intensities, such as materials with multi-layer structures, living biological tissues, etc. The system can obtain the backscattered light of samples to quickly generate the 2D cross-section image, 2D profile and 3D perspective of samples, and has the advantages of non-contact, non-damage, high image resolution and simple operation. At the same time, we also use average, four-phase cancellation and smooth step function to reduce noise, realize the measurement of finger epidermis and dermis and obtain their 3D perspective view. Sweat gland structures were observed in the dermis of the fingers. We also realized the non-destructive measurement of the kapoton tapes, photographed its 2D profile, and obtained its single channel waveform diagram on this basis, and completed the non-destructive and non-contact measurement of the multi-layer structure.
\end{abstract}


\begin{highlights}
\item An optical coherence tomography system was designed to obtain a 3D perspective view of objects with different scattered light intensities.
\item Realized a non-destructive method for acquiring 2D profile images of samples with multi-layered structures\cite{RN32}.
\item Demonstrated a precise calibration method for piezoelectric ceramic displacement stages using a Michelson interferometer and circular truncation algorithm.
\end{highlights}


\maketitle

\section{Introduction}

\par 

With the advancement of science, modern medical imaging technology has played a crucial role in medical diagnostics. Various detection methods and display techniques are becoming more precise, intuitive, and sophisticated, aiding in the observation of biological tissues and comprehension of material structures.

Starting from the discovery of X-rays, doctors have been able to directly observe biological tissues through the skin using various methods. This advancement has significantly elevated the field of medicine. Currently mature medical imaging technologies include X-ray imaging\cite{RN29}, ultrasound imaging\cite{RN31}, fluorescence imaging\cite{RN39}, and magnetic resonance imaging\cite{RN40}. However, these biomedical imaging technologies have certain limitations. For instance, X-ray computed tomography involves relatively high radiation exposure to living organisms, ultrasound imaging has lower resolution, and magnetic resonance imaging requires lengthy scan times, limiting real-time observations. Additionally, due to differences in their underlying physical principles, these technologies exhibit variations in imaging depth and resolution. Techniques like X-ray and ultrasound imaging are commonly used for imaging larger areas with overall resolutions at the millimeter level. Optical coherence tomography (OCT) supplements some of the existing technologies by addressing specific voids in imaging depth. Its superior resolution and non-invasive nature have garnered significant attention. After years of development, OCT technology has been widely used in medical imaging fields, such as ophthalmology\cite{RN43}, dentistry\cite{RN44}, blood vessel\cite{RN45} and stomach\cite{RN46}.

In 1991, the Fujimoto group from MIT initially proposed OCT technology. They primarily used time-domain detection-based point detectors and mechanically scanned reference arms in the OCT system, known as Time-Domain OCT (TD-OCT). TD-OCT resembles a Michelson interferometer \cite{RN24} and operates as a low-coherence system\cite{RN33}. Broadband light sources with low coherence are commonly used in TD-OCT systems. Incident light is split into two beams after passing through a fiber coupler: one entering the reference arm, directed via a collimating lens to a plane mirror movable along the optical axis, serving as the reference beam. The optical delay line\cite{RN46} drives the motion of the reflecting mirror, altering the optical path length of the reference arm. The other beam enters the sample arm and, directed by a lens system, focuses on a specific layer of the sample. The backscattered signal light from various depths of the sample generates interference with the reference beam reflected by the plane mirror in the fiber coupler. These interference lights contain both signal light carrying internal sample information and some optical noise, necessitating pre-phase modulation and post-demodulation for enhanced signal-to-noise ratio.

In 1995, Fercher et al. from the University of Vienna introduced a spectral interference measurement technique to measure intraocular distances\cite{RN20}. The spectrometer, composed of a diffraction grating and a linear array CCD camera, captured interference spectral signals, thus termed Spectral-Domain OCT (SD-OCT). Following the emergence of Spectral-Domain OCT systems, to distinguish from the OCT proposed in 1991, the latter was retroactively termed Time-Domain OCT. 

In 1997, Chinn and Swanson employed grating-tuned external cavity semiconductor lasers as light sources to perform two-dimensional optical coherence tomography imaging on a layered glass structure, demonstrating system resolution and depth by imaging the glass layer structure\cite{RN22}. Systems with a light source having wavelength variations over time were named Swept-Source OCT due to the frequency tuning of the light source.

In 1998, Beaurepair et al. proposed a system, termed Full-Field OCT (FF-OCT), based on the principles of coherent gating, capable of producing a direct two-dimensional frontal view of the sample without scanning\cite{RN27}. The FF-OCT system differed significantly from other OCT systems by utilizing a planar array camera, eliminating the need for scanning to obtain a two-dimensional interference image of the sample at once. The FF-OCT system initially acquires a two-dimensional interference image in the X-Y direction (perpendicular to the optical axis). Then, by changing the position of the reference mirror to phase-shift, it decodes the two-dimensional structural information of the sample. Finally, by manually or using a displacement platform, it alters the axial position of the sample to interfere with different depths within the sample, eventually decoding the three-dimensional structural image. Unlike other OCT systems, which first obtain an image of the sample in the X-Z direction (along the Z-axis of the optical axis) and then perform a lateral scan to obtain the three-dimensional structure, the main advantage of the FF-OCT system is the ability to immediately obtain the two-dimensional interference structure of the sample.  This not only allows observation of the sample's interference before image acquisition but also accelerates imaging speed.

To avoid the impact of light reflection from the optical components on imaging quality in OCT systems, Boccara et al. proposed a novel dark-field fiber interferometer design and practically implemented it into an FF-OCT system, achieving improved imaging results in 2015\cite{RN14}. This approach aims to use an opaque aperture (or block) positioned at the pupil plane of the objective to block specular reflections.

During the process of OCT detection, mismatches between the scattered light in the sample arm and the reflected light in the reference arm can lead to a decrease in the contrast of the resulting images. Addressing this issue, Auksorius etal. proposed a novel high-throughput dark-field fiber interferometer design, which they then applied practically to an FF-OCT (Full-Field Optical Coherence Tomography) system, resulting in improved imaging outcomes in 2020\cite{RN1}.Their design effectively utilizes incident light while suppressing strong reflections within the optical path. Unlike typical FF-OCT systems, their design incorporates an asymmetric beam splitter, achieving a 90\%:10\% split ratio. This asymmetric beam splitter directs the stronger beam toward the test sample and the weaker beam toward the reference mirror. This setup allows for the coherent superposition of two reflections with similar intensities, which are captured by the photodetector. Consequently, this greatly enhances photon utilization and enables high-precision imaging using lower-power light sources.

However, the intensity of the light scattered by different samples varies greatly, the 9:1 split ratio is not widely applicable for various objects. In this work, we try to introduce neutral density filters into the optical path to achieve more matched intensities for the two interference arms. Kapoton tapes with strong scattered light and finger with weaker scattered light were all successfully scanned to complete their 2D profile and 3D reconstruction.

\section{Theory}
\subsection{Characteristics of Broadband Light}
\begin{figure}[htbp]
    \centering
    \includegraphics[width=\linewidth]{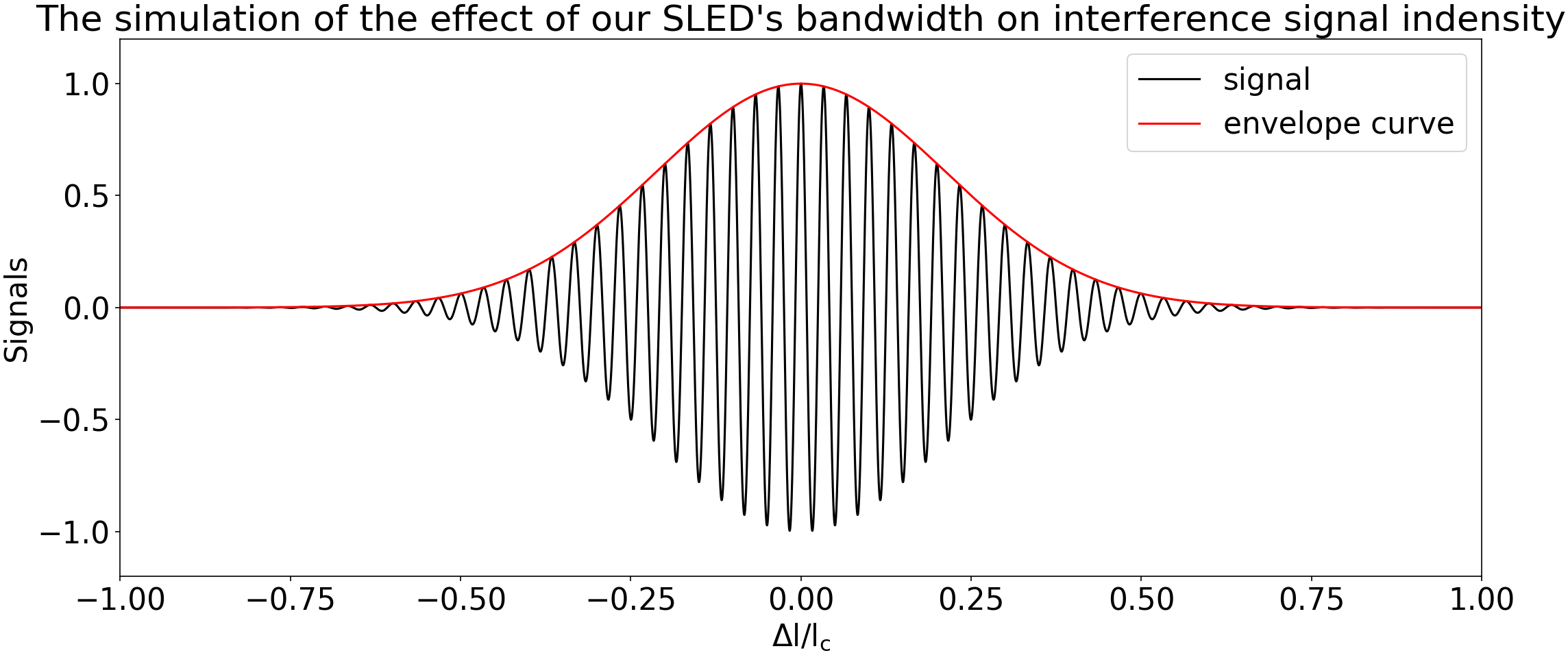}
    \caption{Characteristics of Broadband Light}
    \label{fig:bbl}
\end{figure}

Interference of light is an important characteristic of its wave nature. For two light beams to interfere, several conditions must be met: firstly, both beams must have the same direction of oscillation; secondly, they must have the same oscillation frequency; furthermore, a constant phase difference between the two light beams is necessary; finally, the optical path difference between the two beams must be smaller than the coherence length of the light waves. In optical coherence tomography imaging, interference signals are generated only when the optical path length of the reference light returning and the signal light reflected or scattered from the sample arm are equal or when their optical path difference is within the coherence length of the light source. The diagram \ref{fig:bbl} illustrates the coherent properties of broadband light.
Narrowband light has higher coherence and longer coherence length, whereas broadband light has lower coherence and a shorter coherence length. When narrowband light is chosen as the light source for imaging, the contrast of interference fringes remains relatively constant with changes in the optical path length between the reference and sample arms. However, if broadband light is chosen, its shorter coherence length causes significant changes in fringe contrast with variations in optical path length. Outside the coherence length, interference signals cannot be obtained, enabling interference imaging at specific depths.

\subsection{Coherent length}
The coherent length ($l_c$) of light is a spatial range along the direction of light propagation where the electric field exhibits significant correlation. It is related to the coherence time ($\tau_c$), where $l_c = c \times \tau_c$, with c representing the speed of light. In a steady state, where statistical properties do not change over time, $\tau_c$ is defined as the full width at half maximum (FWHM) of the auto-correlation function $G(\tau)$ of the electric field E(t):
\begin{equation}
    G(\tau)=\int_{-\infty}^{\infty}E(t)E(t+\tau)dt
\end{equation}

According to the Wiener-Khinchin theorem, for a given spectral shape, both the coherence length and coherence time are inversely proportional to the frequency bandwidth.
\begin{equation}
    \int_{-\infty}^{+\infty}G(\tau)\exp(i\omega\tau)d\tau=|E(\omega)|^2=S(\omega)
\end{equation}

where  $S(\omega)$  is  the  power  spectral  density  distribution  of  the  light. 

If  $S(\omega)$  is  Gaussian, 
\begin{equation}
    S(\omega)=\frac{1}{\sqrt{2\pi}\sigma_\omega}\exp\left(-\frac{(\omega-\omega_0)^2}{2\sigma_\omega^2}\right),
\end{equation}

In which, $\omega_0$ represents the central angular frequency, $\sigma_\omega$ represents the standard deviation of $\omega$. As the profile is a crucial focal point, $S(\omega)$ is normalized to unit power.
\begin{equation}
    \int_{-\infty}^{\infty}S(\omega)d\omega=1.
\end{equation}

This gives an expression for the coherence length($l_c$)\cite{RN23}
\begin{equation}
    l_{c}=\frac{2\ln2}{\pi}\frac{\lambda_{0}^{2}}{\Delta\lambda}.
\end{equation}

In which, $\lambda_0$ represents the central wavelength of the light source, $\Delta \lambda$ represents the full width at half maximum (FWHM) bandwidth on the wavelength. The broader the bandwidth, the shorter the coherence length. In an interferometer, when $2n\Delta l < l_c$, the two light beams are considered coherent with each other. n is the refractive index of the medium. $\Delta l$ is the distance difference between sample arm and reference arm.

\subsection{Four-phase cancellation\cite{RN7}}
During each cycle, the reference mirror moves four times with a phase shift of $\varphi=\frac{\pi}{2}$each time. The change in optical path difference is $\frac{\lambda_0}{4}$, and the total distance moved by the reference mirror is $\frac{\lambda_0}{8}$.

Hence, the signal continuously received by the CCD while moving the reference arm can be represented as:

\begin{equation}
    \begin{aligned}
        &I_n\left(x,y\right)\\=&I_0\left(x,y\right)+A\bigl(x,y\bigr)\cos\biggl[\varphi\bigl(x,y\bigr)+\frac{\pi}{2}\bigl(n-1\bigr)\biggr]\quad \\&n=1,2,3,4
        \end{aligned}
\end{equation}

$I_0(x,y)$ represents the DC term, originating from the incoherent light of the sample arm and reference arm. $A(x,y)$ denotes the interference signal amplitude generated at point $P(x,y)$ by the backscattered light from the sample and the reference light reflected from the reference mirror, which constitutes the desired image-depth signal. $\varphi(x,y)$ represents the phase difference produced between the two arms. To extract the tomographic signal $A(x,y)$, it is necessary to eliminate the DC term $I_0(x,y)$.
\begin{figure}[htbp]
    \centering
    \includegraphics[scale=0.1]{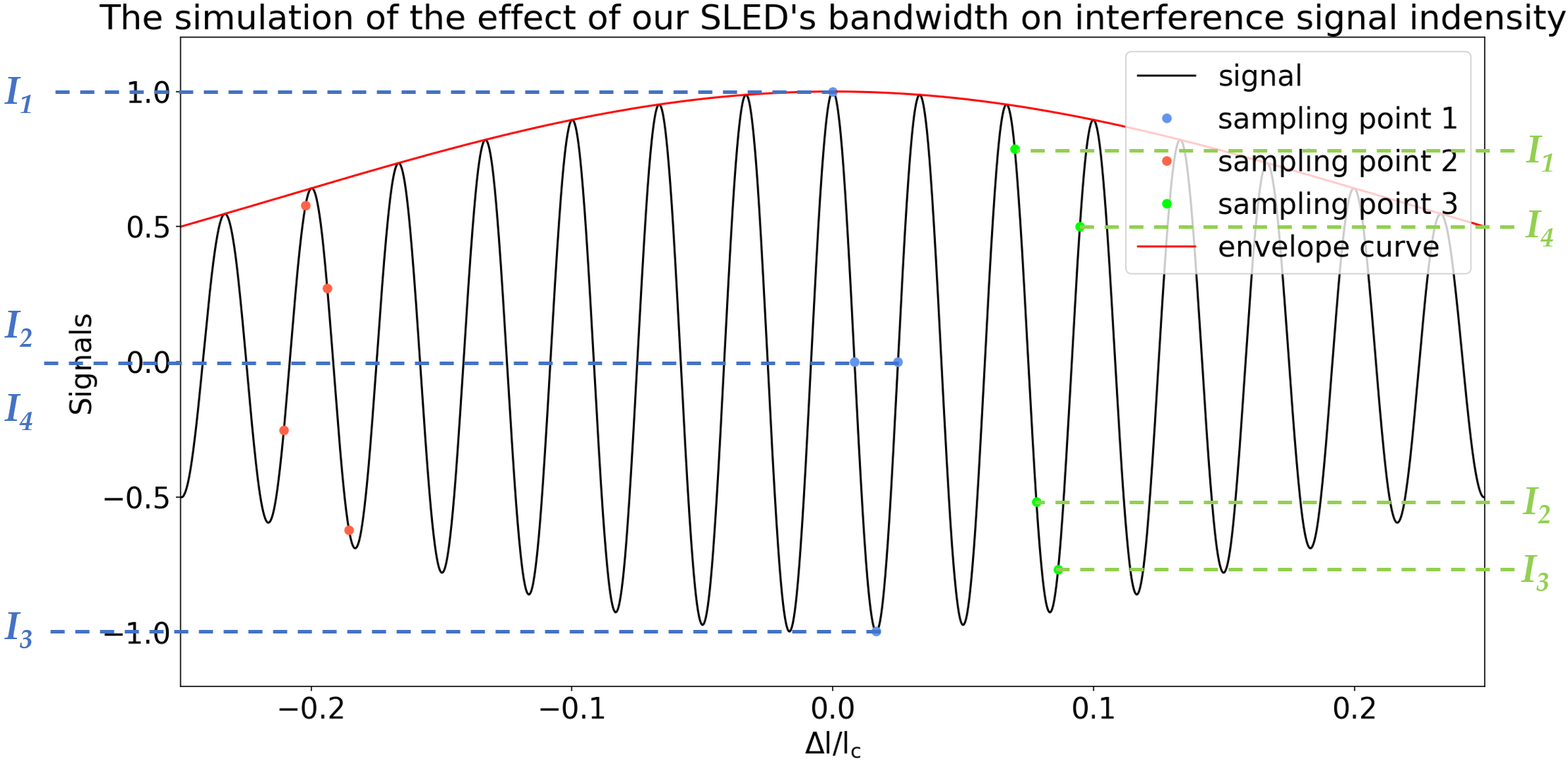}
    \caption{$I_1, I_2, I_3, I_4$ respectively represent the intensity of the AC term corresponding to $n=1, 2, 3, 4$ .}
    \label{fig:4xiangwei}
\end{figure}

Therefore, the resulting tomographic signal at point $P(x,y)$ is

        \begin{equation}
    \begin{aligned}
        &A\big(x,y\big)\\=&\Bigg[\frac{\big(I_1-I_4+I_2-I_3\big)^2+\big(I_1+I_4-I_2-I_3\big)^2}{8}\Bigg]^{\frac{1}{2}}
        \\\sim& \sqrt{(I_1-I_3)^2+(I_2-I_4)^2}
    \end{aligned}
\end{equation}

$n$ represents different positions of the reference arm during phase shifting. $I_1,I_2,I_3,I_4$respectively denote the phase shifts of the reference arm, representing the signals received by the CCD at $0,\frac{\pi}{2},\pi,\frac{3\pi}{2}$.

\section{The experimental setup}
\subsection{High-throughput dark-field FF-OCT setup}
\begin{figure}[htbp]
    \centering
    \includegraphics[scale=0.11]{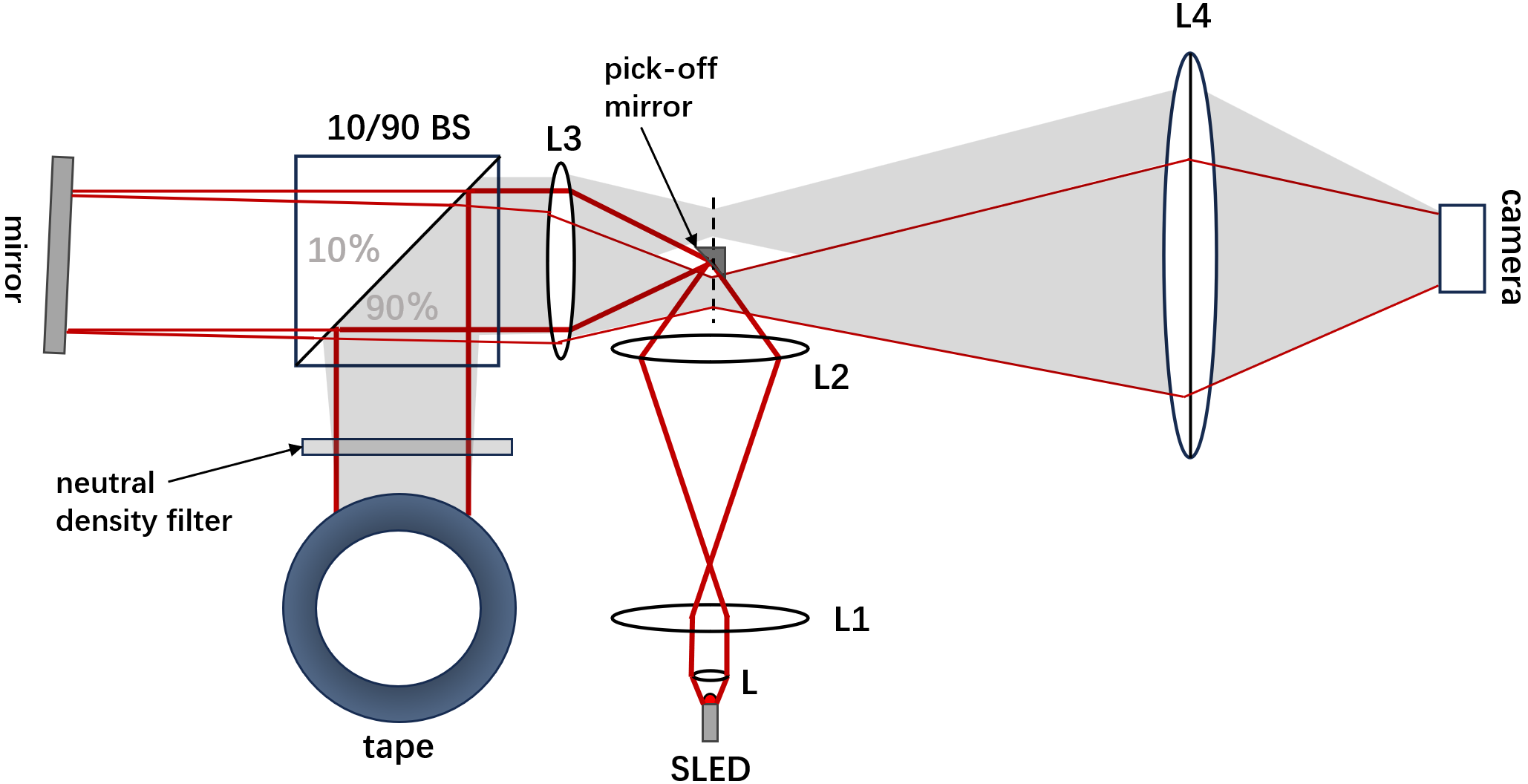}
    \caption{Experimental optical path diagram for filming tape}
    \label{fig:d}
\end{figure}

\begin{figure}[htbp]
    \centering
    \includegraphics[scale=0.11]{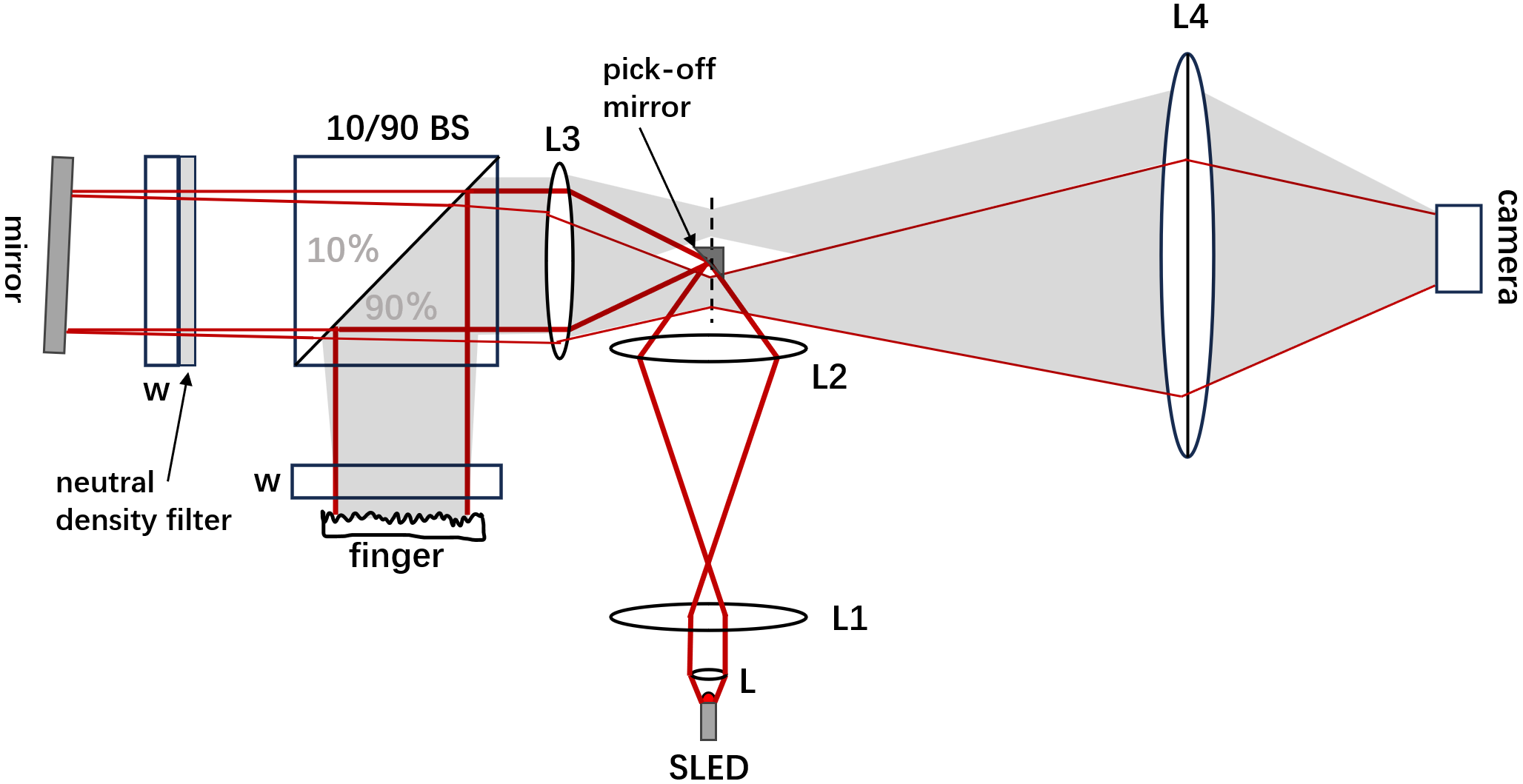}
    \caption{Experimental optical path diagram for filming finger}
    \label{fig:e}
\end{figure}

Figure \ref{fig:d} and Figure \ref{fig:e} show the structure of the HTDF FF-OCT setup used in this experiment. The system utilizes a Superluminescent Diode\cite{RN34} (SLD850, Thorlabs) as the light source, with a central wavelength of 852.2nm and a spectral bandwidth of 50.1 nm. The emitted light, after collimation by lens L, passes through lenses L1 (f=4cm) and L2 (f=4cm) and converges onto a 45° pick-up mirror (Thorlabs). The pick-up mirror, attached to a fine support rod, is positioned at the waist of the Gaussian beam after the reflected light from the reference arm passes through L3 (f=4cm).
A 90/10 beamsplitter (BS) reflects 90$\%$ of the light onto the sample, while the remaining 10$\%$ illuminates the reference mirror after passing through the BS. Most of the backscattered light from the sample is reflected by the BS and imaged onto a CMOS camera (MV-CA013-A0UM, Hikirobot) through L3 and L4, interfering with the reference light reaching the camera.
Lens assembly L4 (f=3.6cm) comprises two plano-convex lenses (f=6cm each). When filming the adhesive tape (as shown in Figure \ref{fig:d}), no damaging action was performed on the tape. Instead, an unopened roll of tape was placed on the stage in front of the BS. Additionally, a neutral density filter was placed in front of the tape to enhance the contrast of interference patterns.

When capturing images of the finger (as depicted in Figure \ref{fig:e}), the finger was pressed against a window mirror coated with an anti-reflection film for 620-1100 nm wavelengths (w). A neutral density filter (Thorlabs) was added to the reference arm to enhance interference contrast.
Due to the use of a pick-up mirror in the optical path, most mirror-reflected light from optical elements is blocked by the pick-up mirror (i.e., directed back to the light source) and not received by the camera (i.e., dark-field detection), thus further improving the system's signal-to-noise ratio compared to traditional OCT systems.
A similar window mirror was inserted into the reference arm to match spectral dispersion between the two arms. Finally, the experimental system imaged a circular sample area with a diameter of 0.6cm on the camera. As the reference mirror is tilted, the image exhibits a slight angular deviation from the imaging collection plane. However, this tilt does not notably reduce the contrast of the fringes along the optical axis, and the entire reference mirror lies within the camera's focal depth range.

In Figures \ref{fig:d} and \ref{fig:e}, the red-colored beams represent the illumination and reference beams, while the grey-colored beams represent the scattered light.

\subsection{Resolution of HTDF FF-OCT}
\subsubsection{Axial resolution}
Using the formula for coherence length:
\begin{equation}
\Delta L=\frac{2\ln{2}}{\pi}\frac{\lambda^2}{|\Delta \lambda|} 
\end{equation}
Where $\lambda$ represents the central wavelength of the broadband light, and $\Delta \lambda$ denotes the wavelength bandwidth of the broadband light. Substituting the basic parameters of the light source: $\lambda=852.2  nm$ and $\Delta \lambda=50.1 nm$, the coherence length is calculated as $\Delta L=6.40 \mu m$. This indicates a noticeable interference contrast within a range of approximately $6.40 \mu m$.

\subsubsection{Lateral resolution}
The lateral and axial resolutions of an OCT system are independent. Therefore, during the design of an optical system, the lateral resolution can be tailored according to specific requirements without affecting the system's axial resolution. The optimal lateral resolution of the system can be determined by the Abbe criterion:
\begin{equation}
\Delta x=\frac{4\lambda_0f}{\pi d}=0.61\frac{\lambda_0}{N.A.}
\end{equation}
Here, $f$ represents the focal length, $d$ denotes the aperture size, and N.A. stands for the numerical aperture. When specific parameters are substituted, the lateral resolution of the lens is calculated as $4.33 \mu m$.

The system's lateral resolution is further constrained by the CMOS camera. The pixel pitch of the CMOS camera used in this experiment is $l=4.80\mu m$, and the lateral magnification factor is $k=0.333$. When this information is substituted into the following equation:
\begin{equation}
    \Delta x=l/k
\end{equation}
It is calculated that due to the restriction imposed by the CMOS pixel density, the lateral resolution is 
$14.4\mu m$. 

The final overall lateral resolution of the system is actually influenced by both of the factors mentioned above. However, theoretically, there isn't a quantitative formula to characterize the functional relationship between the ultimate lateral resolution of the system and the two calculated lateral resolutions mentioned earlier. Nonetheless, in this system, compared to the aperture, the pixel size is relatively large. Consequently, the final lateral resolution of the system is more significantly constrained by the pixel size. Upon capturing the final image, when the image is magnified, visible pixel contours are evident even before reaching the optical resolution limit. This observation indicates that, in practical terms, the ultimate lateral resolution of the system is more dominantly constrained by the size of the pixels.

\subsection{Instrument Calibration}
In this experiment, a uniaxial flexible displacement platform (utilized for nanoscale displacement) is installed on a linear displacement platform (used for micron-level displacement), allowing the platform to achieve both minute nanoscale and larger micron-scale movements simultaneously. However, due to the displacement platform's output being solely in the form of voltage signals, calibration of the displacement platform is required to establish the mapping relationship between voltage and displacement distance.

\begin{figure}[h]
    \centering
    \includegraphics[scale=0.14]{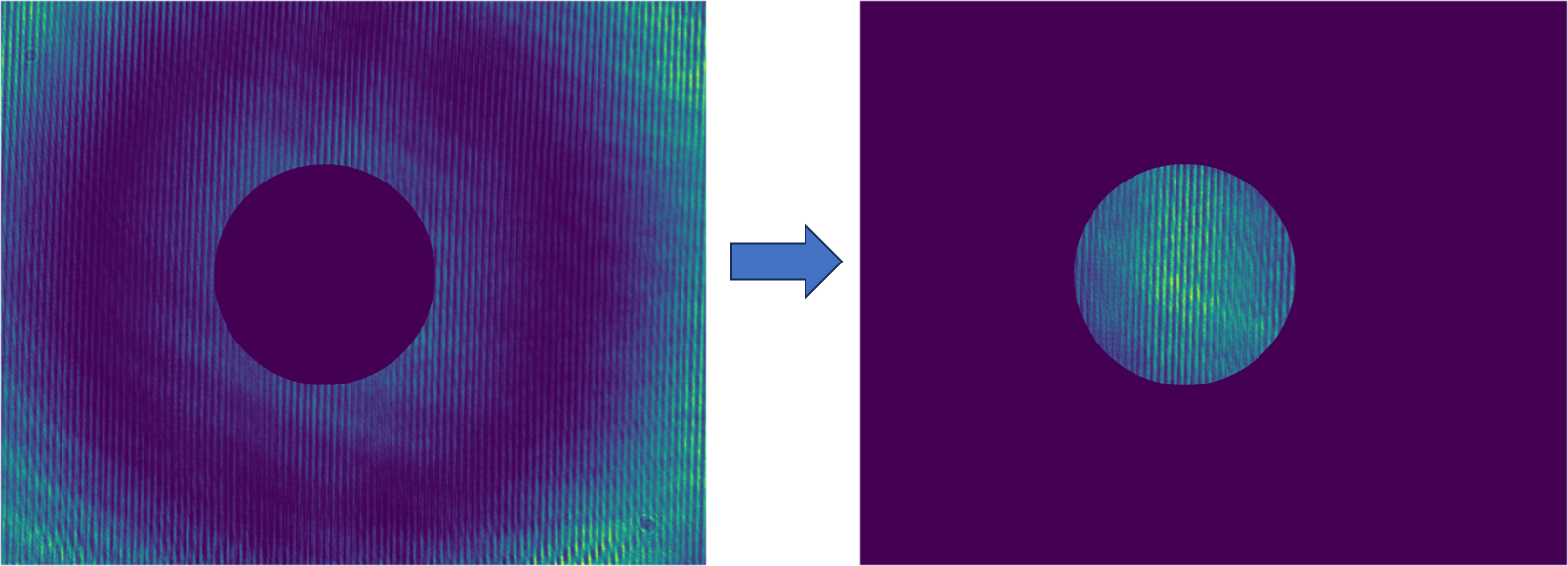}
    \caption{Truncate a Circular Region from the Two-Dimensional Matrix Digital Signal}
    \label{fig:i}
\end{figure}

The experiment utilizes a Michelson interferometry  setup to calibrate the piezoelectric ceramic displacement platform, employing a helium-neon laser (632.8nm) as the light source. However, traditional methods of stripe movement counting are abandoned in favor of a circular truncation method. This involves intercepting a circular section from the 2D matrix digital signal of the interference fringe collected by the CMOS sensor (as shown in Figure \ref{fig:i}). The total intensity of light is computed by summing the intensities of each pixel within the circular region.

\begin{figure}[h]
    \centering
    \includegraphics[width=\linewidth]{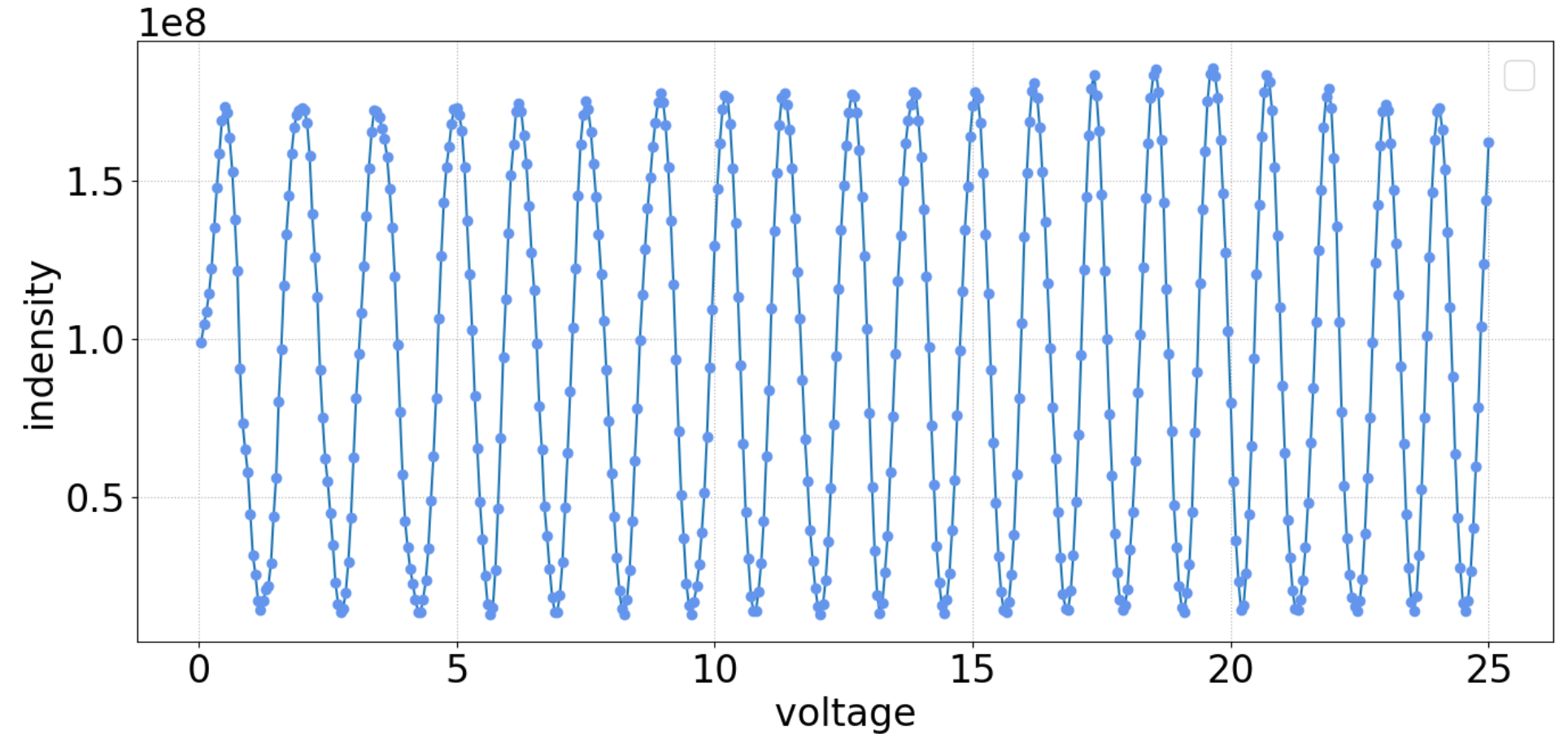}
    \caption{Relationship Graph between Light Intensity and Voltage Variation}
    \label{fig:g}
 \end{figure}

As the voltage of the piezoelectric ceramic changes monotonically, the interference fringe pattern undergoes periodic alterations. Correspondingly, the total light intensity within the circular section also undergoes periodic changes. (One full cycle of total light intensity variation corresponds to a change in optical path difference equivalent to one wavelength, resulting in the displacement platform moving half a wavelength.) The experimental results are illustrated in Figure \ref{fig:g}.

\begin{figure}[htbp]
    \centering
    \includegraphics[width=\linewidth]{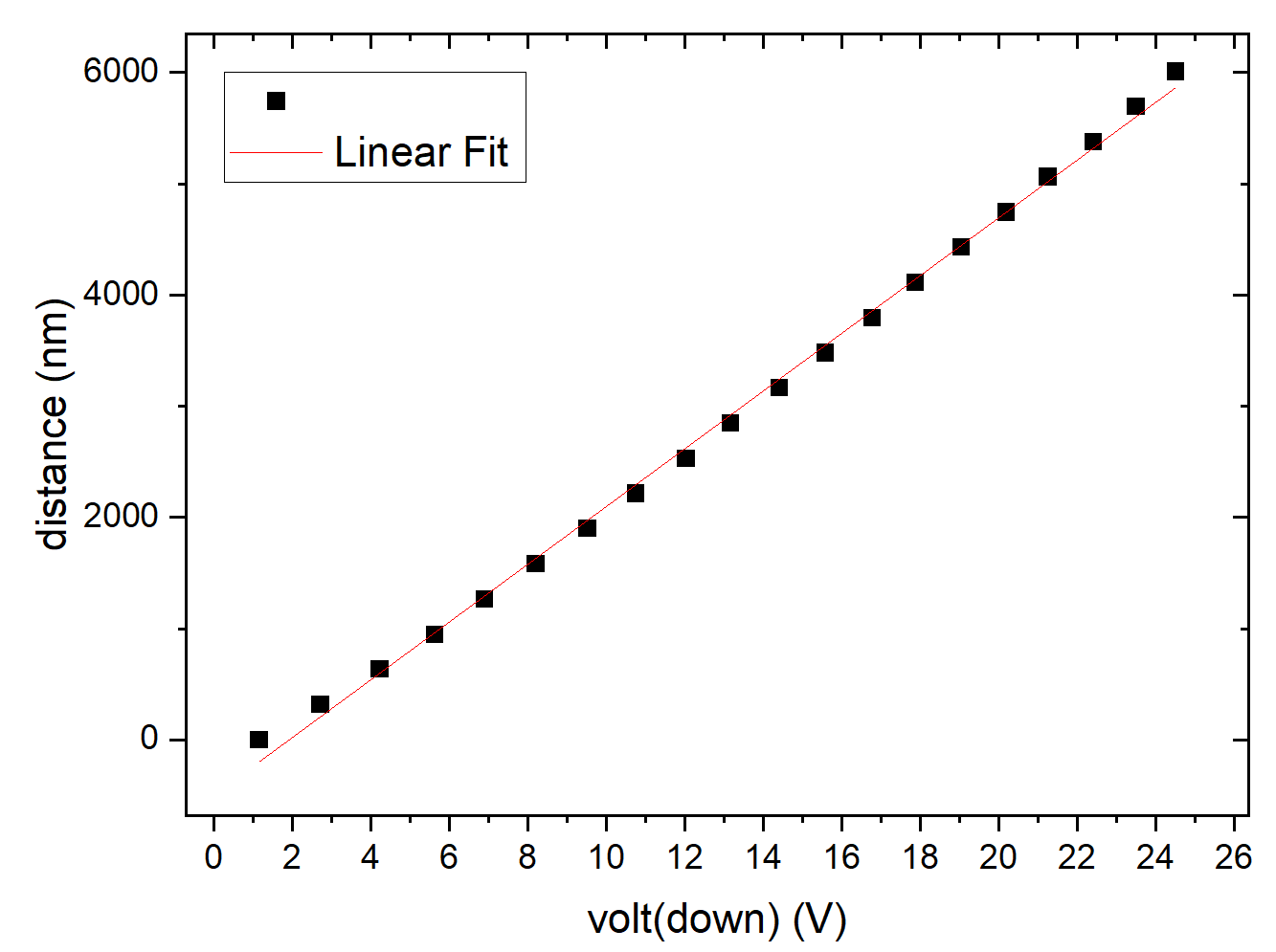}
    \caption{Voltage to Displacement Mapping Function (r=0.99789)}
    \label{fig:k}
\end{figure}

The periodicity of light intensity variation with voltage, as depicted in Figure \ref{fig:g}, exhibits a distinct pattern. We performed fitting using the trough values and obtained the mapping relationship between voltage and displacement, as illustrated in Figure \ref{fig:k}. We opted to fit the trough values because through extensive experimentation, we observed that troughs demonstrate a more "sharpened" characteristic compared to peaks.

The mapping function is defined as $y(nm) = 259.7(nm/V) \cdot x(V) - 499.0(nm)$. Within the range of 8V to 18V illustrated in Figure \ref{fig:k}, the mapping function exhibits the most linearity. Therefore, when manipulating the piezoelectric ceramic displacement platform, the voltage is constrained within this range. The relative error of this method is 1$\%$.

\subsection{Signal Processing}
We set the displacement step of the displacement stage controlled by a stepping motor to ten micrometers and covered a relatively extensive range of total displacement length. At each position, we captured images across four phases, acquiring ten images for each phase. Subsequently, we processed and analyzed the obtained images to determine the range generating coherent signals. Then, we adjusted the step size of the motor-controlled displacement stage to a few micrometers, ensuring that the total displacement distance only covered the range with coherent signals. We repeated the aforementioned measurement procedure to achieve more precise measurement results. Finally, we performed averaging on the obtained images for noise reduction. We employed phase subtraction processing for the four phases at each position to generate two-dimensional cross-sectional images.

\begin{figure}[h]
    \centering
    \includegraphics[width=\linewidth]{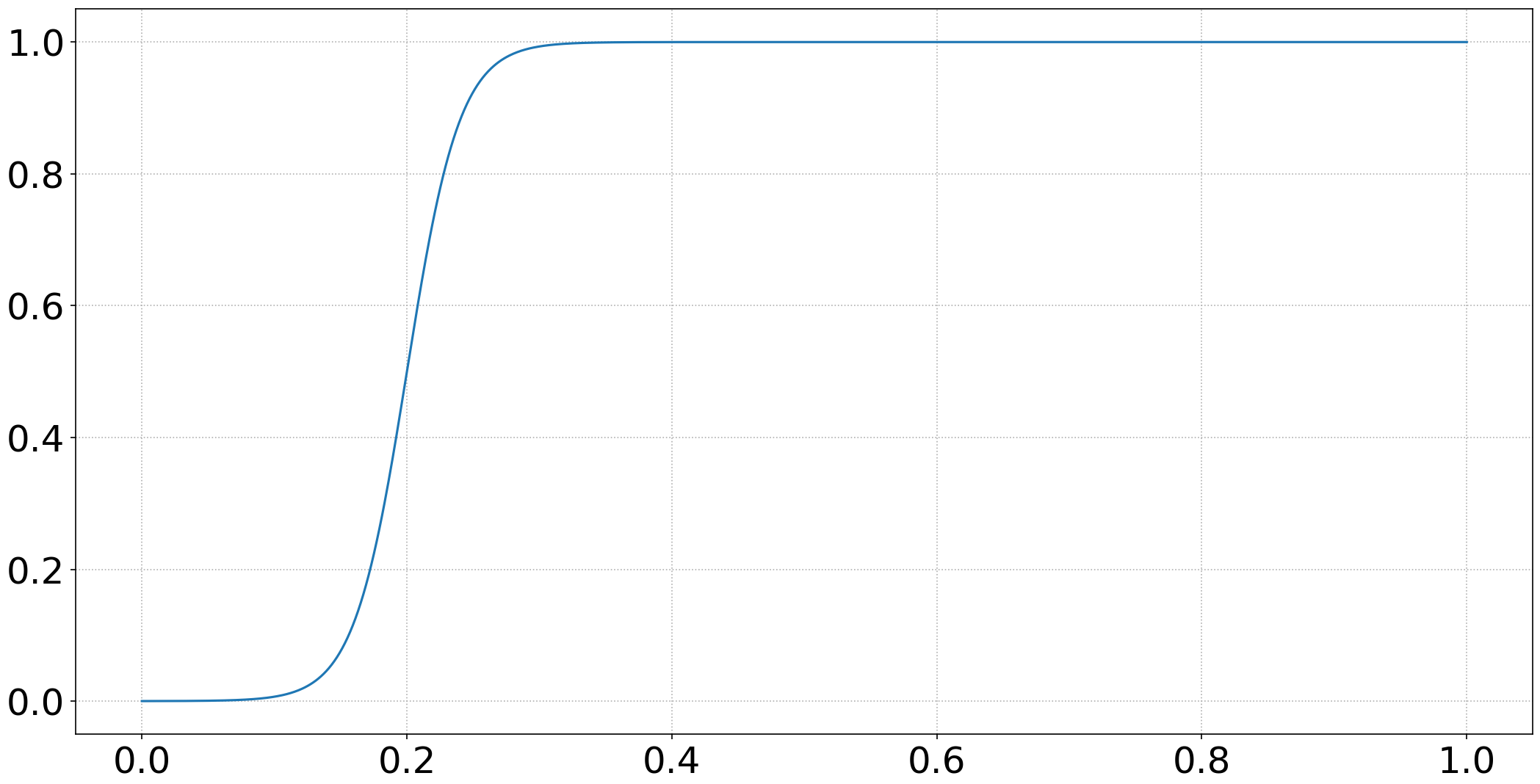}
    \caption{$f(x)=\frac{1}{1+e^{-k(x-x_{0})}}$, $k$ represents the smoothness of the filtering step function, while $x_0$ denotes the position of the step in the filtering step function. The step position is selected based on the intensity of the signal in the incoherent region.}
    \label{fig:u}
\end{figure}

While processing the cross-sectional images, we noticed that the difference in intensity between the useful signal and the residual incoherent signal was not sufficiently distinct. Hence, we devised a smooth step function (Figure \ref{fig:u}) to achieve a better separation between the coherent and incoherent signals.

\subsection{Two-dimensional Profile and Single Channel
Waveform Diagram of Tape}
\begin{figure}[h]
    \centering
    \includegraphics[scale=0.35]{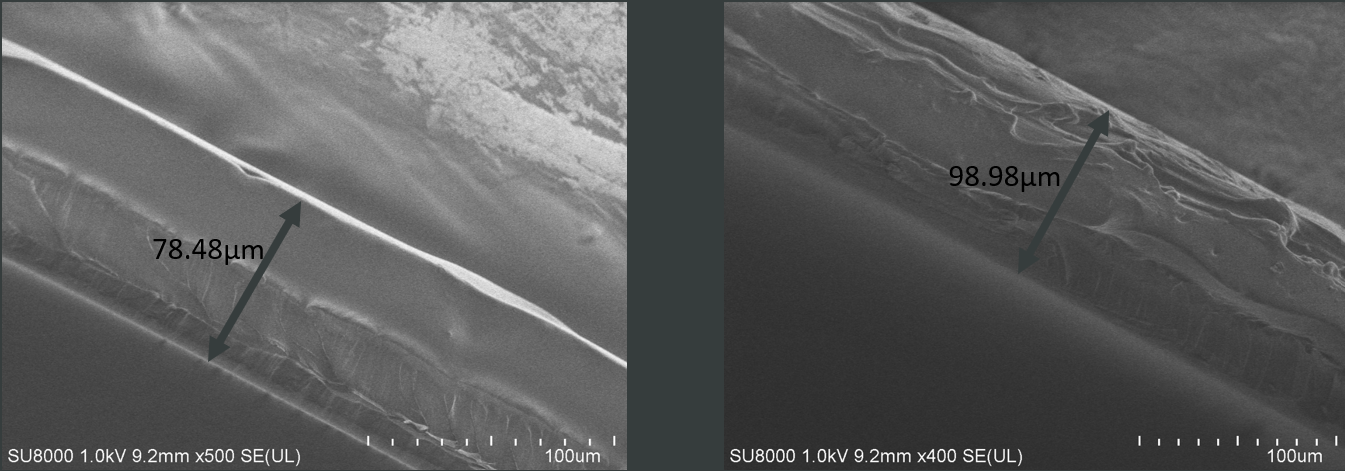}
    \caption{The Profile of Tape under SEM}
    \label{fig:dj}
\end{figure}

The tape is a typical translucent material with a multi-layered structure, making it an ideal candidate for demonstrating the OCT system. Typically, a single-layer tape comprises a substrate and an adhesive layer adhered to it. For this demonstration, we have chosen two types of high-temperature insulation tapes with thicknesses of 80 $\mu m$ and 100 $\mu m$ (selected based on detailed specifications in their datasheets, high material purity, and minimal thickness variance). The 80 $\mu m$ tape consists of a substrate layer thickness of 50 $\mu m$ and an adhesive layer thickness of 30 $\mu m$. Meanwhile, the 100 $\mu m$ tape comprises a substrate layer thickness of 75 $\mu m$ and an adhesive layer thickness of 25 $\mu m$. Figure \ref{fig:dj} shows an 
scanning electron microscope image of the tape (the undulations on the profile are remnants from cutting the tape to fit into the scanning electron microscope).

\begin{figure}[h]
    \centering
    \includegraphics[scale=0.5]{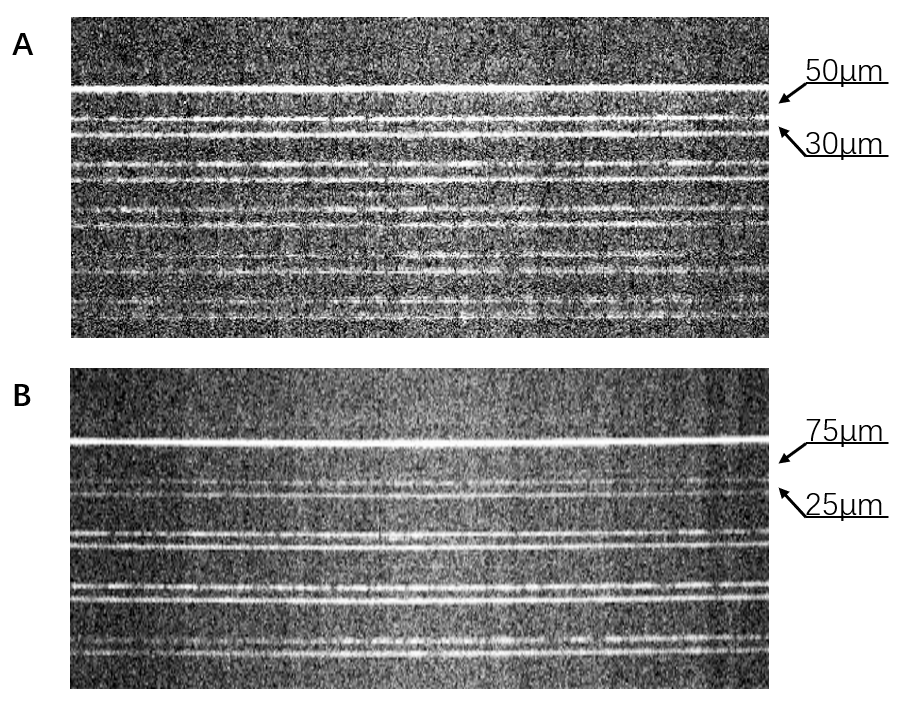}
    \caption{(A) Tape with a total thickness of 80 $\mu m$, with a substrate thickness to adhesive thickness ratio of 5:3.
    (B) Tape with a total thickness of 100 $\mu m$, with a substrate thickness to adhesive thickness ratio of 3:1.}
    \label{fig:r}
\end{figure}

Using our OCT system, we can measure cross-sectional images at different positions of the tape without damaging it. These cross-sectional images can then be stacked along the axial direction to form a 3D representation\cite{RN35}\cite{RN37}. Subsequently, a longitudinal section is extracted to obtain a two-dimensional profile image of the tape, as shown in Figure \ref{fig:r} (left: tape with a thickness of 80 $\mu m$, right: tape with a thickness of 100 $\mu m$).

\begin{figure}[htbp]
    \centering
    \includegraphics[width=\linewidth]{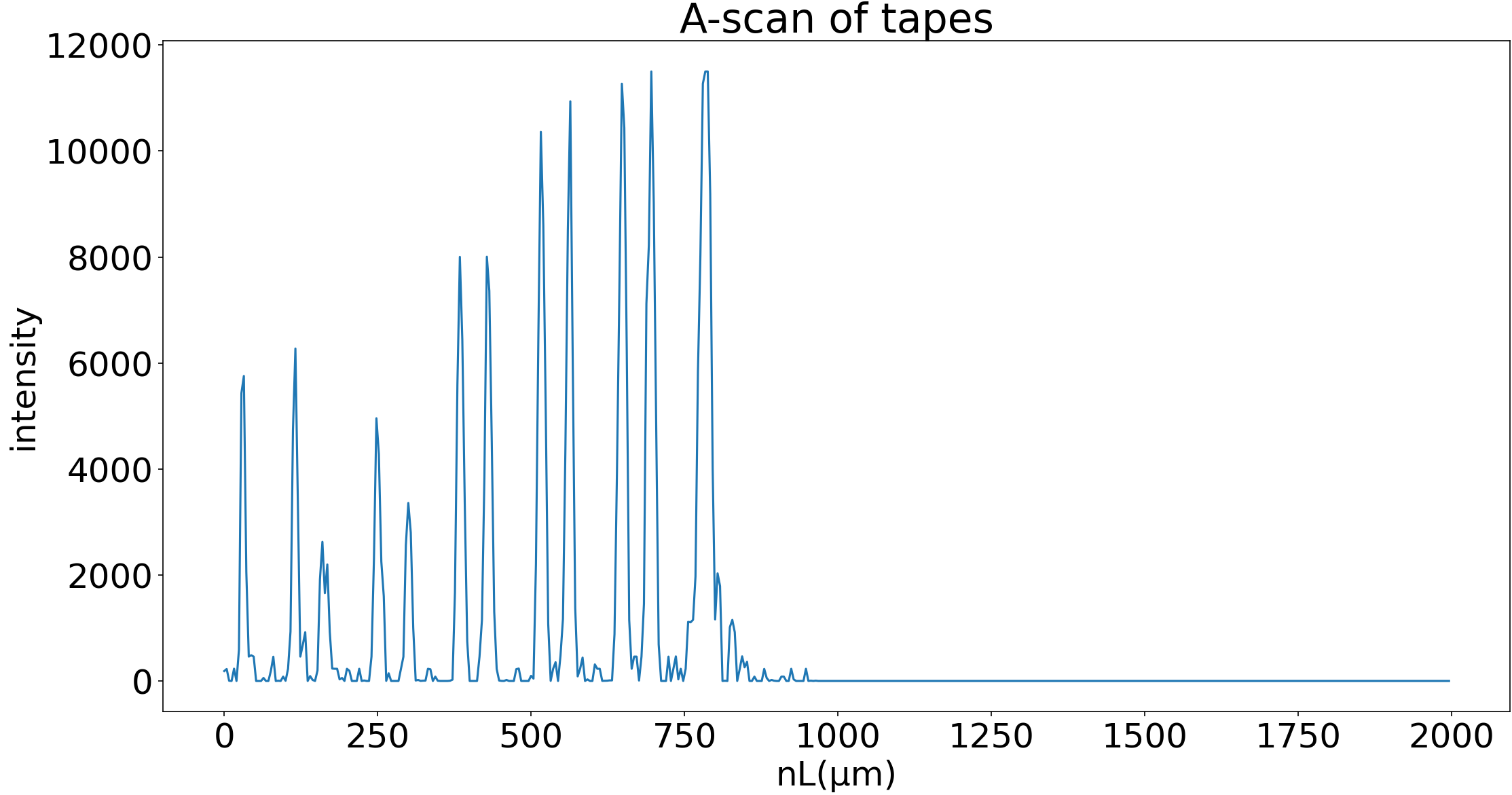}
    \caption{Single Channel Waveform of the 80 $\mu m$ Tape}
    \label{fig:80}
\end{figure}

\begin{figure}[htbp]
    \centering
    \includegraphics[width=\linewidth]{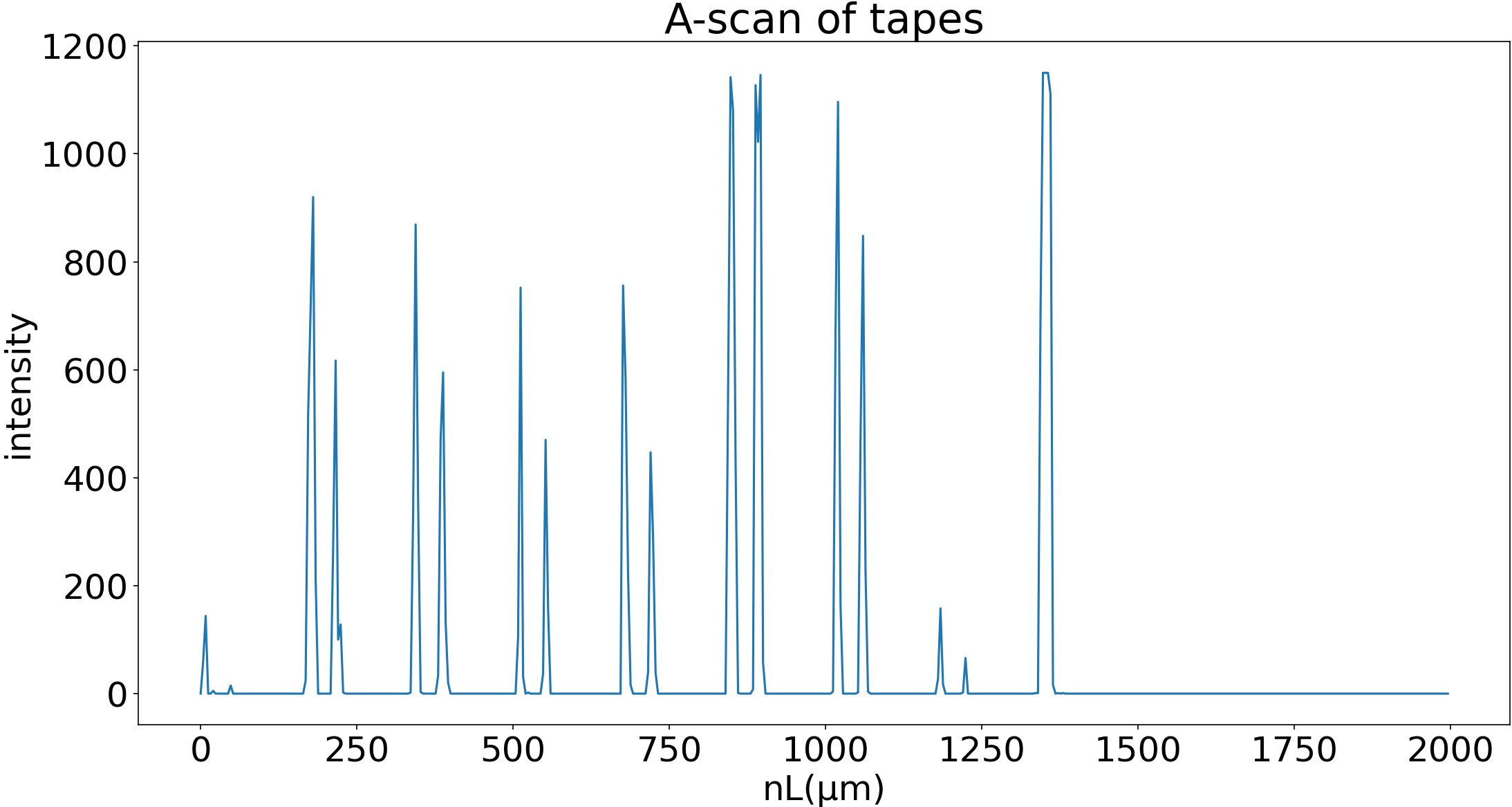}
    \caption{Single Channel Waveform of the 100 $\mu m$ Tape}
    \label{fig:100}
\end{figure}

Next, we extracted individual waveforms from the two-dimensional profile images by taking horizontal line profiles, resulting in Figure \ref{fig:80} and Figure \ref{fig:100}. In these figures, the scattering signals at interlayer interfaces are notably visible, and the 'peaks' have a half-width at approximately the micron level, indicating excellent axial resolution of the system. By analyzing the peak value differences in the individual waveforms and considering the material of the tape (polyimide) with a refractive index approximately 1.6, thickness data for different layers of the tape can be obtained. The results of the data analysis are presented in Table \ref{table:2}. Reading the half-height width of a-scan of the tape from Figure \ref{fig:80}and Figure \ref{fig:100}, the actual longitudinal resolution can be obtained as 10.94 $\mu m$ which is better than 14$\mu m$ in the reference\cite{RN1}.

\FloatBarrier
\begin{table*}[htbp]
    \caption{Adhesive Tape Measurement Results}
    
    \label{table:2}
    \begin{tabular*}{\linewidth}{@{} LLLLL@{} }
        \toprule
        & \multicolumn{2}{c}{\textbf{80$\mu m$tape}} & \multicolumn{2}{c}{\textbf{100$\mu m$tape}} \\
        \cmidrule(lr){2-3} \cmidrule(lr){4-5}
        & Optical Path Length($\mu m$) & Thickness($\mu m$) & Optical Path Length($\mu m$) & Thickness($\mu m$) \\
        \midrule  
        S1$(\mu m)$ & 84.05 & 52.53 & 119.2 & 74.50 \\
        A1$(\mu m)$ & 43.97 & 27.48 & 46.80 & 29.25 \\
        S2$(\mu m)$ & 88.04 & 55.03 & 122.8 & 76.75 \\
        A2$(\mu m)$ & 51.95 & 32.47 & 42.80 & 26.75 \\
        S3$(\mu m)$ & 84.01 & 52.51 & 129.6 & 81.00 \\
        A3$(\mu m)$ & 43.95 & 27.47 & 40.00 & 25.00 \\
        S4$(\mu m)$ & 88.08 & 55.05 & 126.0 & 78.75 \\
        A4$(\mu m)$ & 47.94 & 29.96 & 43.20 & 27.00 \\
        S5$(\mu m)$ & 84.01 & 52.51 & 127.6 & 79.75 \\
        A5$(\mu m)$ & 48.02 & 30.01 & 43.20 & 27.00 \\
        \midrule 
        Average(S)$(\mu m)$ & \multicolumn{2}{c}{53.53} & \multicolumn{2}{c}{78.15} \\
        Average(A)$(\mu m)$ & \multicolumn{2}{c}{29.48} & \multicolumn{2}{c}{27.00} \\
        \bottomrule
    \end{tabular*}
   S1,S2,S3,S4,S5 are substrate layer of tape.
A1,A2,A3,A4,A5 are adhesive layer of tape.
\end{table*}

\FloatBarrier
\subsection{finger}
\begin{figure}[htbp]
    \centering
    \includegraphics[width=\linewidth]{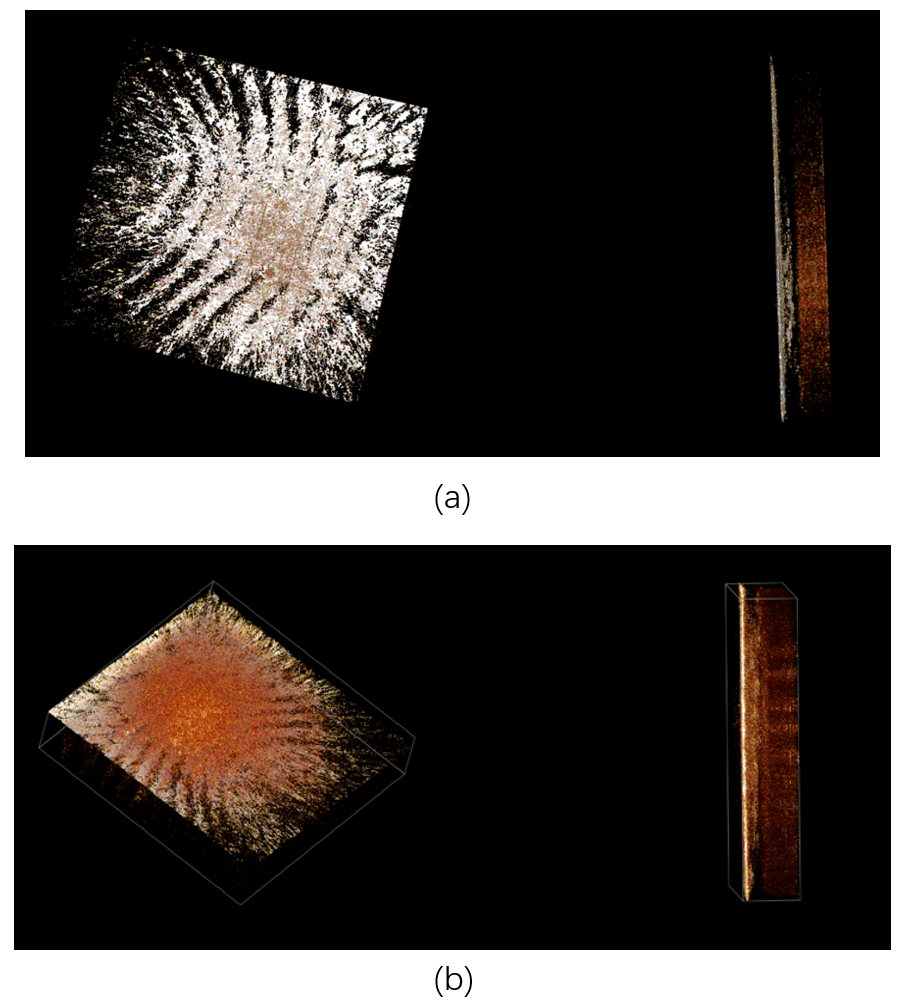}
    \caption{(a) The image (a) is a 3D reconstruction composed of 100 sectional images stacked with a spacing of 8mm. In this image, the signal from the dermal layer is weak, and there is significant edge attenuation. Therefore, we normalized the epidermal and dermal layers separately, setting the peak value of the epidermal layer to 1 and the peak value of the dermal layer to 0.8, resulting in image (b). (b) According to image (b), the boundary between the epidermal and dermal layers is evident, and subcutaneous sweat glands can be observed.}
    \label{fig:4figer}
\end{figure}

Our OCT system has a penetration depth of 0.5mm for fingerprint imaging, capturing signals from the epidermal layer and partial dermal layer. We collected fingerprint cross-sectional images at various depths using the OCT system and stacked them axially using ImageJ, resulting in the reconstruction of the fingerprint. This process produced a three-dimensional perspective view of the fingerprint(Figure \ref{fig:4figer}).
\FloatBarrier

\begin{figure}[htbp]
    \centering
    \includegraphics[width=\linewidth]{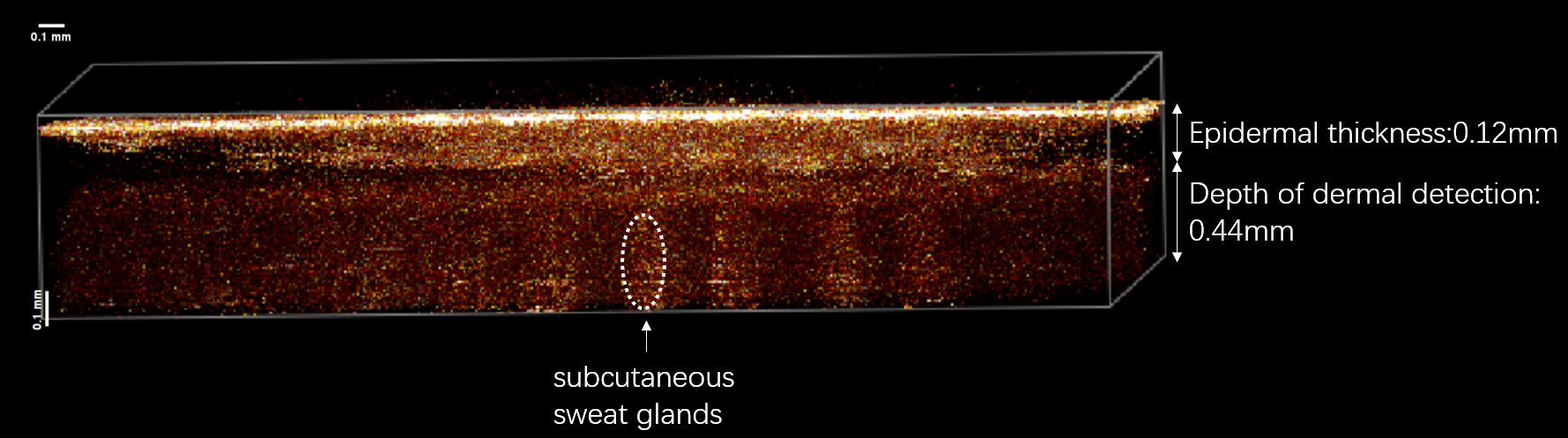}
    \caption{Side view of a fingerprint}
    \label{fig:fin}
\end{figure}

\FloatBarrier
Figure \ref{fig:fin} shows the detail of the side view of a fingerprint. Analyzing the image and considering the refractive index of a finger to be around 1.4, it can be concluded that the epidermal layer thickness of a finger's fingerprint is 0.12 mm, and the probing depth of the dermal layer is 0.44 mm. Sweat gland structures can be observed in the dermis of the fingers, and one of them is circled with a dotted line to guide observation. 

Hence, this instrumentation system exhibits the capability for high-resolution imaging of the epidermal and dermal layers, enabling non-invasive detection and diagnosis of various skin lesions, including epidermal and dermal pathologies. This facilitates improved early diagnosis and treatment of skin disorders. Moreover, the micron-level resolution of this instrument system enables clear visualization of the superficial microvascular structures within the skin, including capillaries and microvascular networks, which can be utilized to assess the vascular distribution and density in different skin lesions. For instance, when detecting inflammatory skin\cite{RN47} conditions or tumors\cite{RN49}, it can be used to observe changes in vascular density.

\section{Conclusion}
This paper is based on the fundamental principles of high-throughput dark-field full-field optical coherence tomography. It comprehensively demonstrates the establishment of a measurement system capable of examining various samples with different scattered light intensities. Measurements were conducted on both tapes and fingerprints using this FF-OCT system. Utilizing this system, we implemented noise reduction techniques such as averaging, four-phase phase cancellation, and a smooth step function. This facilitated measurements of the epidermal and dermal layers of finger skin, generating their 3D perspective views. Simultaneously, we performed non-destructive measurements on tape, capturing its 2D profile images and obtaining single channel waveform diagrams. This allowed for non-destructive, non-contact measurements of multi-layered structures. Furthermore, we presented a precise calibration method for piezoelectric ceramic displacement stages using a Michelson interferometer and circular truncation algorithm.
\newpage

\bibliographystyle{unsrt}  

\bibliography{htdfOCT}

\end{document}